# High-Fidelity Hole Spin Qubits Reveal Quadrupolar Nuclear-Bath Dynamics in Isotopically Purified Planar Germanium

Jian Zeng[1,2,3,7], Xiangjun Tan[1,4,7], Hongzhang Wang[1], Yulei Zhang[1,6], Wendong Bian[1], Lingting Wu[1,2,3], Chenggang Yang[1,2,3], Zhengshan Guo[1], Jiankun Li[1], Yongfeng Wang[6], Jun Lu[5*], Jun-Wei Luo[3,5], Tian Pei[1*]

[1]*Beijing Academy of Quantum Information Sciences, Beijing 100193, China*
[2]*Beijing National Laboratory for Condensed Matter Physics, Institute of Physics, Chinese Academy of Sciences, Beijing 100190, China*
[3]*University of Chinese Academy of Sciences, Beijing 100049, China*
[4]*Department of Physics and Astronomy, University College London, WC1E 6BT London, United Kingdom*
[5]*State Key Laboratory of Semiconductor Physics and Chip Technologies, Institute of Semiconductor, Chinese Academy of Sciences, Beijing 100083, China*
[6]*School of Electronics, Peking University, Beijing 1000871, China*

[7]These authors contributed equally: Jian Zeng, Xiangjun Tan
*Corresponding author: peitian@baqis.ac.cn and lujun@semi.ac.cn

**Abstract**

Planar Germanium has emerged as a promising platform to build spin-based large scale quantum computers. By exploiting the anisotropic hyperfine interaction of holes in Ge, qubits with long $T_2^*$ have been recently realized. While the performance of single qubits is still more or less limited by $^{73}$Ge nuclear spin fluctuations, the site-to-site variation of qubit sweet spot becomes obstacles to maintaining high fidelity of each qubit across the whole wafer. To achieve high performance Ge-based quantum circuit, it is therefore essential to eliminate the origin source of hyperfine noise. In its Silicon counterparts, reduction of $^{29}$Si abundance enables exceptional high-fidelity operation. In contrast, hole qubits based on isotopically purified Ge have not been demonstrated. Here, we report the synthesis of high quality 2-dimensional hole gas (2DHG) with enriched $^{70}GeH_4$ precursor. Due to the suppression of nonzero spin nucleus, the qubits' $T_2^*$ on the sweet spot is moderately extended beyond 20 μs, surpassing the previous best reported Ge hole qubits. More importantly, the qubits' $T_2^*$ off the sweet spot is enhanced to above 3 μs, enabling single qubit gate fidelity exceeding 99.9% in both operating regimes. Hahn-echo spectroscopy further resolves a finite-frequency nuclear-noise channel that is distinct from the conventional Larmor-linked hyperfine response. We associate this channel with quadrupole-modified dynamics of residual $^{73}$Ge nuclei sampling local electric-field gradients near the Ge/SiGe interface. Its field scaling and angle-dependent visibility are consistent with a qubit-visible quadrupolar nuclear-noise component transduced through the anisotropic hyperfine interaction of Ge holes. These results establish isotopically purified planar Ge as a high-coherence scalable platform for hole spin qubits and provide a spectroscopic probe of interfacial quadrupolar nuclear dynamics.

Semiconductor-based spin qubits represent a leading platform for scalable, fault-tolerant quantum computing[1], offering high-density integration and compatibility with established industrial fabrication processes[2–5]. To achieve long coherence time, Group IV materials are used as the host platforms to minimize hyperfine interaction from non-zero spin nuclei[6–11]. In silicon (Si) system, the performance of electron spin can be further enhanced via isotope purification[12,13]. By reducing the abundance of $^{29}$Si to 800 ppm or lower, $T_2^*$ has been extended to 100 μs scale[14,15], enabling high-fidelity qubit gates[16–20].

While silicon has long been the focus of this field, Germanium (Ge) has recently emerged as a high-performance alternative[21–24]. The valence band of Ge hosts holes with strong intrinsic spin-orbit coupling, which allows for rapid, all-electrical qubit control without the need for external micromagnets[25]. Significant progress has been made in Ge hole qubits, from the demonstration of multi-qubit logic in two-dimensional arrays[26–30] to quantum simulation of many-body interacting systems[31].

Compared with silicon electron qubit, the dipole-dipole hyperfine interaction in Ge 2-dimensional hole gas (2DHG) is intrinsically weak and strongly anisotropic[32–35]. When the qubit quantization axis is aligned to the sweet plane by adjusting external magnetic field orientation, the Ising-type coupling between nuclear spin and hole spin is minimized, extending $T_2^*$ beyond 17.6 μs[36]. However, residual hyperfine interactions persist and the sweet spot for hyperfine interaction does not always coincide with that for charge noise. Since the proposal of spin qubit based on planar Ge system[10,37], it is anticipated that isotopic purification of Ge could significantly reduce hyperfine interactions arising from nuclear spin and extend qubits' coherence time, much like that in Si system. However, there has still been no experimental verification to date. More importantly, spatial variations of g-tensor cause sweet spots for different qubits vary across the wafer, so compromise in the orientation of the uniform external field is needed[27]. This underscores the necessity of isotopic purification to further elevate hole qubit performance and to improve the robustness of Ge based quantum circuits' scaling.

In this work, we demonstrate the synthesis of high-mobility germanium quantum wells by using isotopically purified precursors. These heterostructures enable the definition of high-performance, electrically gated quantum dots that serve as robust hosts for spin qubits. Due to the suppression of $^{73}$Ge abundance, qubit's $T_2^*$ is improved by one order of magnitude when operated off the hyperfine sweet spots compared with the natural Ge devices. Furthermore, by aligning the external magnetic field within the hyperfine sweet spots, we reach a benchmark $T_2^*$ of 22.5 μs, a new record for hole spin qubits. The gate fidelity of single qubit both on and off the hyperfine sweet spot reaches 99.9%, demonstrating that the robustness of Ge hole qubit scalability can be enhanced by isotopic purification. Beyond improving the coherence metrics, the reduced hyperfine background allows Hahn-echo spectroscopy to resolve an additional finite-frequency nuclear-noise component that is not captured by a single Larmor-linked $^{73}$Ge precession branch. Field-dependent echo measurements and modelling separate this response from the conventional Larmor branch and associate it with nuclear quadrupole resonance (NQR), quadrupole-modified dynamics of residual $^{73}$Ge nuclei near the Ge/SiGe interface. In this picture, finite-field Zeeman-quadrupole mixing sets the relevant nuclear eigenstates, while anisotropic hyperfine transduction converts the resulting nuclear-bath dynamics into qubit-frequency noise. The angular dependence of the extracted visibility suggests that the local electric field gradient (EFG) axes, hole quantization axis, and hyperfine-transduction tensor need not coincide. While quadrupole effect has previously been investigated as a decoherence source in Si[38] and GaAs[39–41] spin qubits, and as a modulation tool for nuclear spin transitions of a single nucleus[42], our studies use the hole qubit to probe the dynamics of the nuclear bath in planar Ge, which is modified by the local EFG. Because the EFG is set by heterostructure growth, interface disorder, strain, and gate electrostatics, the measured nuclear response carries local information about the quantum-dot environment and suggests a route to strain and EFG sensing with Ge hole-spin qubits.

## High quality $^{73}$Ge purified 2DHG

The Ge quantum well (QW) heterostructure was epitaxially grown via ultra-high vacuum chemical vapor deposition (UHV-CVD) using a reverse-grading technique[43], with the active well layer deposited specifically from isotopically enriched $^{70}GeH_4$ precursor which is converted from $GeO_2$ having a $^{73}$Ge abundance of 0.195%. As illustrated by the transmission electron microscopy (TEM) image in Fig. 1a, the active region consists of a 14 nm $^{70}$Ge quantum well, sandwiched between natural $Ge_{0.8}Si_{0.2}$ barriers, and protected by a 2 nm Si cap. Secondary ion mass spectrometry (SIMS) analysis in Fig. 1b confirms the high isotopic purity of the quantum well, with $^{73}$Ge concentrations estimated to be 0.4% which is slightly higher than that of the precursor due to $^{73}$Ge residue in the chamber. Notably, the abundance of $^{73}$Ge in the enriched QW is reduced by approximately 20-fold compared to its natural abundance, which is crucial for suppressing hyperfine-induced decoherence. To evaluate the transport quality of the 2DHG, magneto-transport measurements were performed at 1.5 K, revealing well-defined Quantum Hall Effect (QHE) and Shubnikov-de Haas (SdH) oscillations (Fig.1c). Integer Quantum Hall plateaus marked by orange dashed lines and signature of fractional Quantum Hall plateau ($\nu$=4/3) marked by purple dashed line are observed, demonstrating the high quality of the 2DHG. From the Hall resistance and longitudinal resistivity, the hole mobility was determined to be $3.73\times10^5$ cm$^2$/Vs at a carrier density of $1.76\times10^{11}$ cm$^{-2}$. Samples from the same wafer but different fabrication batches yield peak mobility from $3.73\times10^5$ cm$^2$/Vs to $9.30\times10^5$ cm$^2$/Vs (see Extended Data Fig. 1). The density-dependent mobility (Fig. 1d) exhibits two distinct transport regimes: a high-density regime characterized by a weak dependence ($\alpha_H = 0.16$) and a low-density regime with a pronounced power-law dependence ($\alpha_L = 1.46$). While the peak mobility remains below the current records for natural Ge QWs, the transport is primarily limited by background impurities scattering and interface-roughness scattering within the enriched layer, as evidenced by the weak dependence in high-density regime[44].

Achieving the $10^6$ $cm^2/Vs$ regime will necessitate more stringent precursor purification to further suppress oxygen-related background charges, alongside optimized SiGe/Ge interface abruptness to mitigate interface-roughness scattering.

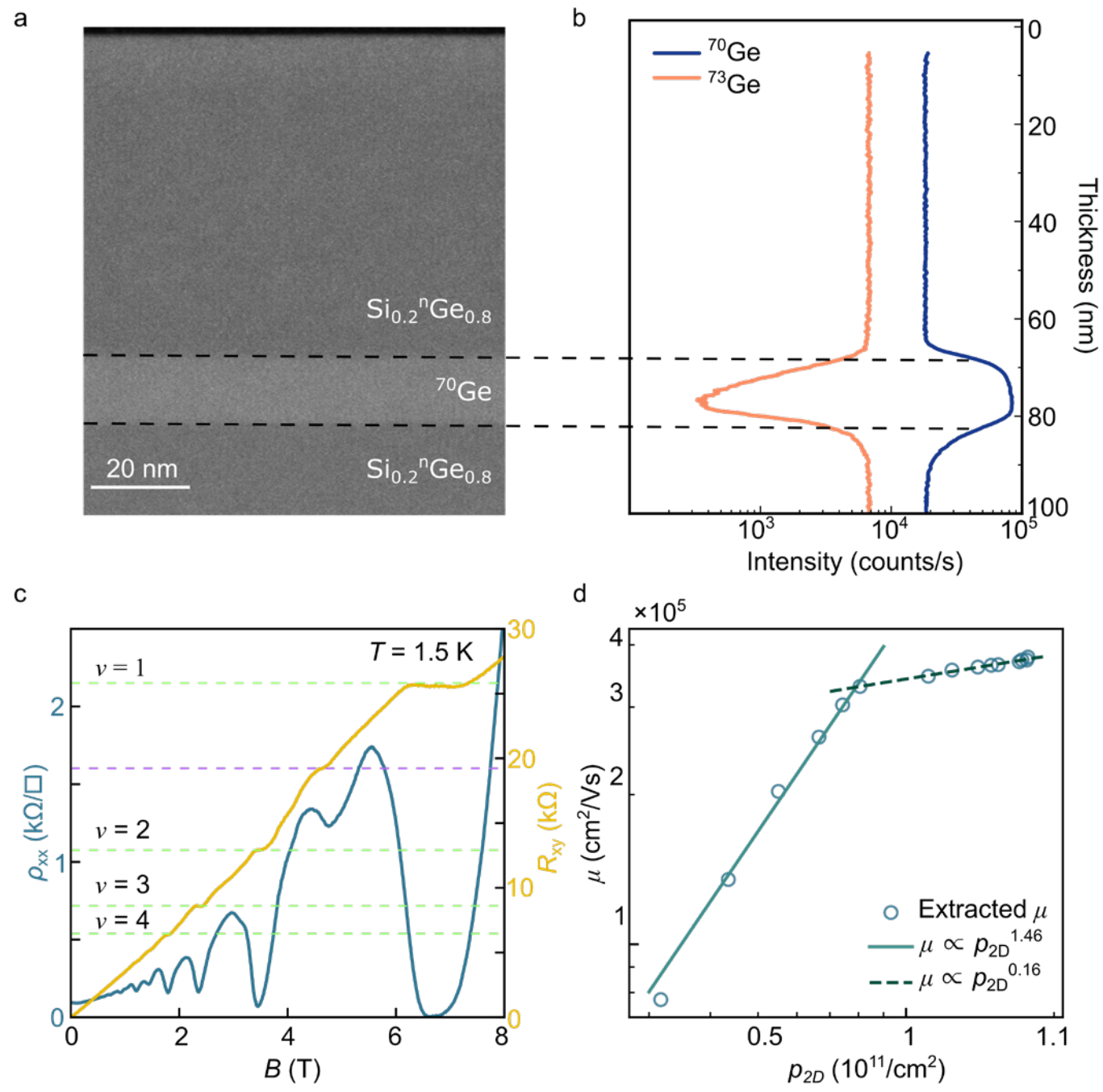

**Fig 1 Wafer Characterization. a,** TEM image of the Ge heterostructure. **b,** SIMS analysis of the heterostructure. The black dashed lines mark where the isotopically purified quantum well is. The tailoring of the intensity is caused by intermixing induced by ion sputtering during testing. **c,** Quantum Hall measurement of the 2DHG. The green dashed lines indicate the integer quantum hall plateaus, and the purple dashed line corresponds to fractional quantum hall filling factor of 4/3. **d,** Density-dependence of the hole mobility. In the low-density regime, mobility shows stronger dependence on density while in the high-density regime, local disorders start dominating.

## Quantum Dot Device based on isotopically purified 2DHG

To construct spin qubits, quantum dot devices are fabricated by standard nano-fabrication techniques. A scanning electron microscopic (SEM) image combined with schematic illustration is shown in Fig. 2a. By tuning the energy levels of the 2DHG via the gate electrodes, a double quantum dot system can be formed (marked by orange circles in Fig. 2a) beneath plunger gates B2 and B4 to host the qubits. An adjacent single hole transistor (SHT) serves as a charge sensor to detect the charge occupancy change in the double dot. A radio frequency tank circuit is connected to one of the accumulation gates of the SHT for high bandwidth readout. Fig. 2b displays the charge stability diagram of the double quantum dot detected by the SHT. By tuning the voltages applied to B2 and B4, the occupancy of the two dots marked by ($N$, $M$) can be depleted to (0, 0), where $N$ ($M$) donates the number of charges occupying the left (right) dot. The tilted vertical addition lines and horizontal addition lines in the stability diagram indicate the adding of a hole into the left dot and the right dot respectively. At more negative gate voltages, the two dots merge into a single dot. The distinct vertical and horizontal addition lines confirm a well-defined double quantum dot system, while their regular spacing reflects the high uniformity of the 2DHG Fermi level and quantum well quality. We investigate the spin qubit in the (1, 1) → (2, 0) region (Fig. 2c) where Pauli Spin Blockade (PSB) can be utilized to achieve high fidelity readout. The kink of the addition line (marked by red dashed line in Fig. 2c) indicates the hole tunnel rate between the right dot and the right reservoir (defined by B5) is much lower than that between the left dot and the left reservoir (defined by B1). This configuration enables a latched readout and improves the relaxation time $T_1$ during the readout phase.[45–48] By applying voltage pulses to B2 and B4 (as shown in Fig. 2d), the qubits undergo an Initialization (I) → Manipulation (M) → Readout (P, L) cycle with each operation point illustrated in Fig 2c. For qubit manipulation, an external magnetic field is applied and a microwave burst is applied to B3 (see Methods). When the driving frequency, $f_{\text{drive}}$, matches with the magnetic field ($hf_{\text{drive}} = \mu_B|\boldsymbol{g} \cdot \boldsymbol{B}|$, where $\boldsymbol{g}$ is the $\boldsymbol{g}$-tensor of the qubit), qubits will be rotated coherently by electrical driven spin resonance (EDSR).

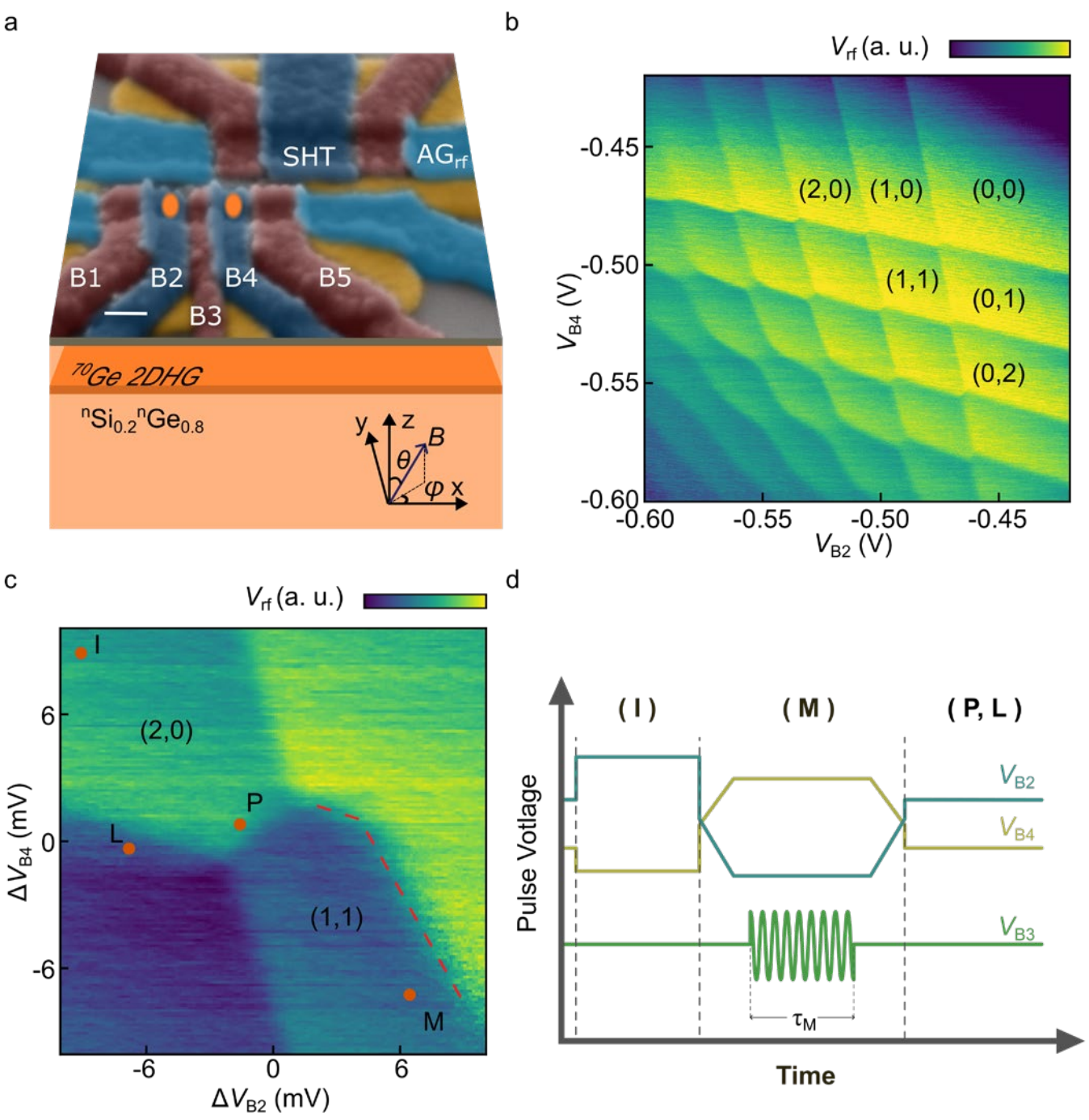

**Fig 2 A double quantum dot device. a**, False-colored scanning electron micrograph of the quantum dot device identical to the one used in this experiment. Double dot is formed beneath plunge gates B2 and B4, respectively. A SHT defined next to the double dot serves as charge sensor, while a tank circuit is connected to $AG_{rf}$ for high bandwidth rf readout. **b,** Charge stability diagram of the double dot detected by the SHT. ($N$, $M$) denotes there are $N$ ($M$) holes in the left (right) dot. The disappearance of the regular spacing addition lines in the upper-right corner indicates that the double dot occupancy is depleted to (0, 0). **c,** Zoom-in charge stability diagram of (2,0) → (1,1) region. Red dashed line marks the kinked addition line because of low tunnel rate between the right dot and the right reservoir. **d,** Schematic diagram of the pulse sequence applied to B2, B3 and B4. The three stages of the pulse correspond to Initialization (I), Manipulation (M) and Readout (including PSB region (P) and Latched readout (L)). The time scale shown in the figure does not represent actual experimental time.

## Qubit Characterization

Figure. 3a shows the Rabi oscillation of one qubit as a function of manipulation time $\tau_M$ and manipulation frequency. Due to the strong anisotropic g-factor and hyperfine interaction in Ge 2DHG[21,32], we measure the EDSR signal as a function of $\theta_B$ around $90^{\circ}$ to identify the hyperfine interaction sweet spots for both qubits. Two parabolic resonance lines are shown in Fig. 3b, indicating that the two qubits possess different g-tensor. Considering the out-of-plane g-factor of Ge hole is much larger than the in-plane g-factor, the hyperfine sweet spots should be around the bottom of the parabolic curves. That is, the hyperfine sweet spots of the two qubits do not coincide with each other. By setting the magnetic field polar angle $\theta_{\boldsymbol{B}} = 90.1^{\circ}$ to tune Q1 into its hyperfine sweet spot, the dephasing time $T_2^*$ is extracted to be 22.5 μs by Ramsey experiment in Fig. 3c. Comparing with previous study where qubits in natural Ge are operated on the hyperfine sweet spot, the $T_2^*$ is moderately improved[36]. This demonstrates that even though the Ising-type hyperfine interaction is already minimized when the external field aligns within the hyperfine sweet plane, the hyperfine interaction persists and the reduction of $^{73}$Ge abundance can further improve the coherence of the qubit. To examine the effect of isotope purification, $T_2^*$ of both qubits are characterized by tilting the magnetic field away from the sweet spots. As shown in Fig. 3d, the sweet spots of the two qubits differ from each other by about $0.5^{\circ}$, consistent with the angle dependent spin resonance spectroscopy. Once the field is rotated away from the sweet spots, the $T_2^*$ decrease rapidly. Nevertheless, in the compromised region, which is about $1^{\circ}$, the $T_2^*$ of both qubits maintain above 5 μs. When the field is rotated away from the sweet spot by about $1^{\circ}$, $T_2^*$ of the qubits degrades to 3 μs scale at $B = 35$ mT. This is about 5 times higher than those $T_2^*$ measured in natural planar Ge/SiGe devices without identifying the hyperfine sweet spot (see Extended Data Fig. 2 and ref. [[10,24,49]]). Given that the $T_2^*$ limited by hyperfine interaction is proportional to $1/\sqrt{N}$, where $N$ is the number of nonzero spin nucleus overlap with the hole's wavefunction[50], the measured $T_2^*$ when qubit operated off the sweet spot in our experiment is consistent with the 20-fold reduction of $^{73}$Ge abundance. By further reducing the $^{73}$Ge abundance to $< 0.01\%$, the lower limit of $T_2^*$ can be improved to beyond 10 μs off the sweet spots, which provides greater robustness for scaling the number of qubits.

To analyse the decoherence of the qubits, we measured $T_2^*$ of both Q1 and Q2 as a function of magnetic field magnitude on Q1's hyperfine sweet spot. As shown in Fig. 3e, by increasing $B$ from 35 mT to 100 mT, $T_2^*$ of Q1 suffers significantly degradation from 22.4 μs to 5.5 μs, while $T_2^*$ of Q2 only decrease mildly from 5.1 μs to 2.7 μs. According to previous study, the charge-noise limited $T_2^*$ is inversely proportional to the Larmor frequency of the qubit. This is in good agreement with the $1/B$ fitting of Q1's $T_2^*$, indicating that for the qubit operated on hyperfine sweet spot, its coherence is mainly limited by charge noise. On the other hand, interaction between nuclear bath and the hole spin is still the major source of decoherence for Q2. While the reason of this noise can be intuitively attributed to the residue $^{73}$Ge in the quantum well and unpurified $^{29}$Si and $^{73}$Ge in the barrier layer, another possible origin will be discussed in the following section.

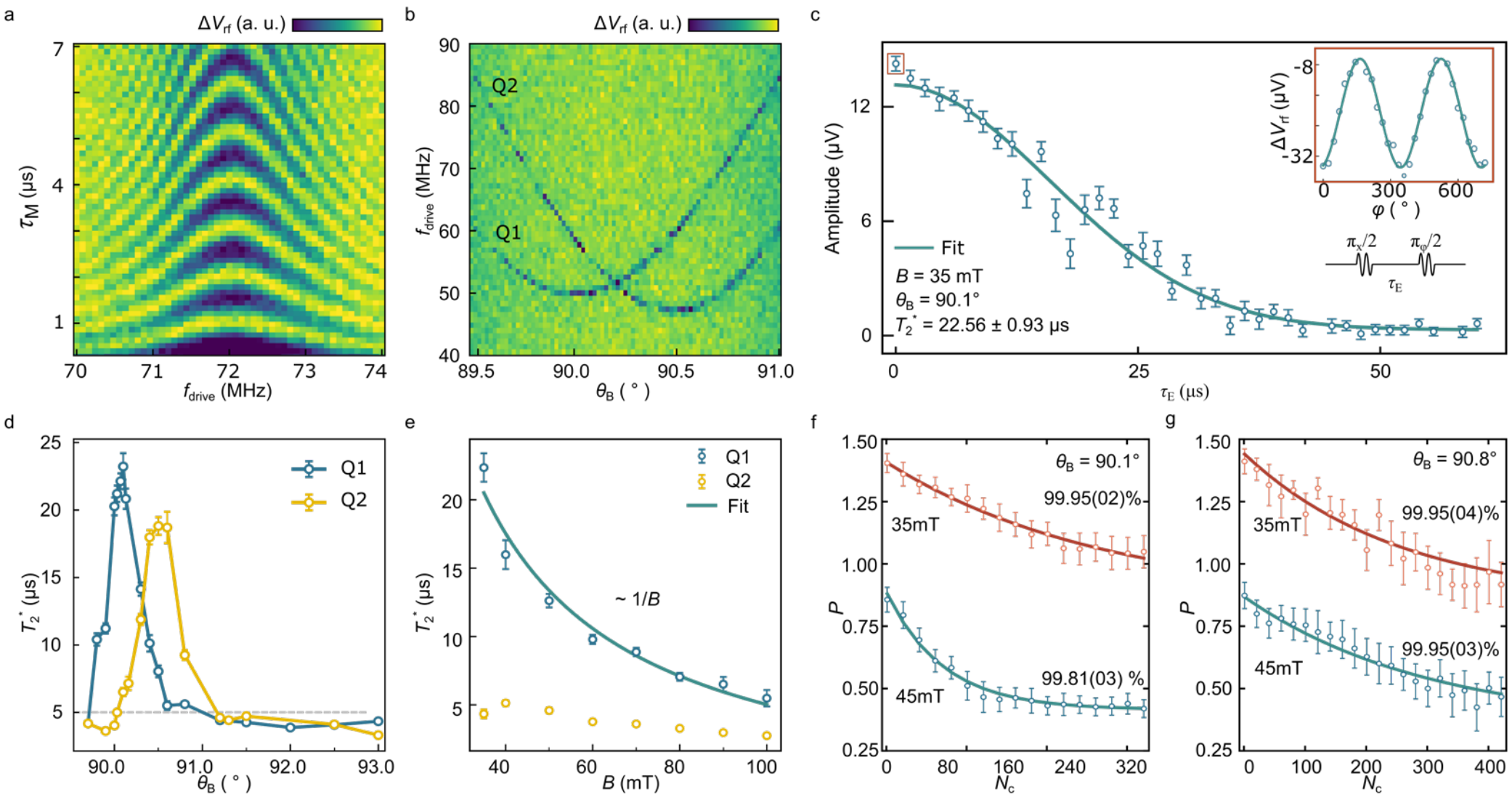

**Fig 3 Coherent manipulation of single qubits. a,** Rabi chevron pattern with an in-plane magnetic field $B$ = 35 mT. **b,** Magnetic field spectroscopy measured at $B$ = 35 mT. Two parabolic resonance lines correspond to the EDSR signals of two qubits, which are labelled as Q1 and Q2. **c,** Dephasing time of Q1 measured by Ramsey experiment at $\theta_B = 90.1^{\circ}$. Points in the figure are amplitude of Ramsey fringes, measured with $\frac{\pi_x}{2} - \tau_E - \frac{\pi_\varphi}{2}$ microwave burst sequence (bottom inset). The upper inset displays the measured signal intensity (dots) as a function of the final microwave burst phase, $\varphi$, at $\tau_E$ = 1 μs, together with cosinusoidal fit (line). The dephasing time $T_2^*$ is obtained by fitting with $A = a\exp[-(\tau_E/T_2^*)^2]+b$. **d,** Dephasing time $T_2^*$ as a function of the magnetic field polar angle $\theta_B$ at $B$ = 35 mT. The optimal angle for each qubit coincides with the minimum point of its parabolic resonance line in Fig. 3c. $T_2^*$ exceeds 20 μs on the sweet spot. When the field rotate away from the sweet spot, $T_2^*$ degrades rapidly and saturates at about 3 μs. **e,** $T_2^*$ of both qubits as a function magnetic field magnitude $B$ with $\theta_B = 90.1^{\circ}$ (Q1 hyperfine sweet spot). $T_2^*$ of Q1 follows a $1/B$ decay, indicating that charge noise dominates decoherence at hyperfine sweet spot. The weaker field dependence of Q2's $T_2^*$ reveal that hyperfine interaction dominates dephasing when qubits operated off the sweet spot. The randomized benchmarking of single qubit (Q1) gate at $\theta_B = 90.1^{\circ}$ (**f**) and $90.8^{\circ}$ (**g**) under magnetic fields of 35 mT and 45 mT respectively. The traces in each figure are offset by 0.5 for clarity. While the qubit achieves a fidelity above 99.9% at $90.8^{\circ}$ (off the sweet spot) at both field magnitudes, its fidelity falls below 99.9% at 45 mT at $90.1^{\circ}$ (on the sweet spot) due to a slower Rabi oscillation frequency. The legend are the extracted average gate fidelities obtained from the Clifford gate fidelities using a ratio of 3.625 (see Methods). The errors are calculated from the 1σ standard error of the mean value for all data fitting.

Qubit performance is evaluated by randomized benchmarking (RB) when it is operated both on and off the hyperfine sweet spots. Average gate fidelity as high as 99.95% is reached when $B$ = 35 mT and $\theta_B = 90.1^{\circ}$, but degrades to 99.81% when $B$ increases to 45 mT (Fig. 3f). When the field orientation deviates slightly to $\theta_B = 90.8^{\circ}$, the average fidelity is 99.95% when $B$ is set to 35 mT and 45 mT (Fig. 3g). This can be explained by how $T_2^*$ and Rabi frequency change as a function of field magnitude. When the qubit is operated on sweet spot, $T_2^*$ decreases rapidly with increasing field magnitude (Fig. 3e) due to charge noise, while Rabi frequency only increases linearly (see Extended Data Fig. 3a) from 296 kHz to 346 kHz. However, when operated off the sweet spot, qubit's $T_2^*$ is mainly limited by hyperfine interaction, so merely reduced by higher field. As a result, the increasing of Rabi frequency (from 371 kHz to 451 kHz) leads to higher fidelity. Moreover, the qubits' Rabi frequency decreases when field rotates from 90.8° to 90.1° (see Extended Data Fig. 3b), this adds extra advantage to performing the qubit off the sweet spot. Nevertheless, qubit operated both on and off the sweet spot can achieve single qubit gate fidelity above 99.9%. Due to the extremely small window of the sweet spot, most qubits in a large circuit are unlikely to be tuned into their sweet spots. Our results not just demonstrate the robustness of the qubit in isotopically purified planar Ge, but also points out that performing the qubit off the sweet spot where higher Rabi frequency is available might be a better choice. This highlights the importance of further reduction of the $^{73}$Ge abundance to obtain qubits with higher $T_2^*$ and quantum circuit with better robustness.

## Quadrupolar $^{73}$Ge nuclear-spin noise in Hahn-echo spectroscopy

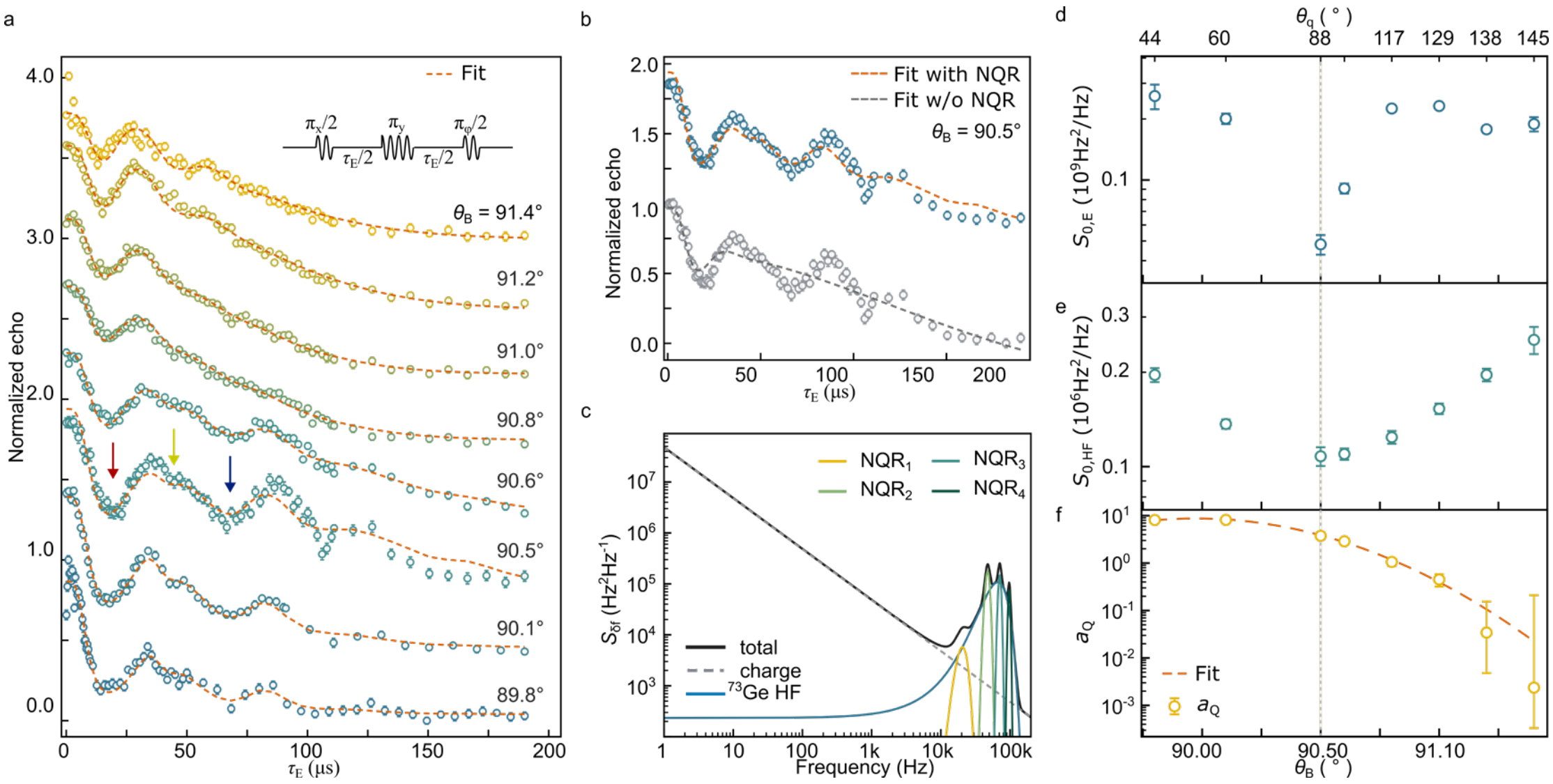

**FIG 4. Quadrupolar nuclear-noise components probed by Hahn-echo spectroscopy. a,** Hahn-echo measurements of Q2 at different magnetic-field polar angles $\theta_B$ with $B = 45$ mT. The extracted echo amplitudes are normalized by the measured maximum-minimum range and vertically offset for clarity. Dashed curves are fits using a noise PSD containing charge noise, a conventional $^{73}$Ge Larmor-linked hyperfine contribution, and ensemble-broadened quadrupolar components. The arrows mark representative Hahn-echo filter maxima rather than independent transition frequencies. **b,** Representative trace at $\theta_B = 90.5°$, comparing fits with and without the quadrupolar component. **c,** Qubit frequency-noise PSD inferred from the fit at $\theta_B = 90.5°$, showing the $1/f$ charge-noise background, the Larmor-linked hyperfine peak, and qubit-visible quadrupolar branches, with the dominant fitted branch centred near 47.3 kHz. **d–f,** Extracted charge-noise prefactor $S_{0,\mathrm{E}}$, hyperfine amplitude $S_{0,\mathrm{HF}}$, and effective quadrupolar visibility $a_\mathrm{Q}$ as a function of $\theta_B$. The hyperfine amplitude is constrained by the local g-factor map and Ramsey $T_2^*$ boundary, while $a_\mathrm{Q}$ shows a pronounced angle dependence consistent with coupling to a local EFG, and the dashed line is a phenomenological signed-angle fit. The magnetic-field direction is converted to the qubit quantization-axis direction (upper axis) based on the measured g-tensor map (see Extended Data Fig. 4).

The decoherence noise is investigated by performing Hahn-echo measurement on Q2. Fig. 4a compares the Hahn-echo decay of Q2 with different $\theta_B$ at $B$ = 45 mT. Characteristic collapse-and-revival features are observed. Conventionally, these are explained by the precession of the nuclear-spin bath about the external magnetic field. Under such a mechanism, the collapse timing scales with the $^{73}$Ge Larmor frequency $\gamma_{Ge-73}$ = 1.49 MHz/T. Indeed, the first collapse dip (red arrow, $\tau_E \approx$ 15-20 μs) matches the expected $^{73}$Ge Larmor period. However, the broad second/third dip (yellow/blue arrow) cannot be explained by higher harmonics of the same precession, nor by $^{29}$Si nuclei ($\gamma_{Si-29}$ = 8.465 MHz/T), which would produce collapse at much shorter times at $\tau_E \approx$ 2.6 μs.

To resolve this discrepancy, we note that $^{73}$Ge is the only stable Ge isotope with nonzero nuclear spin $I = \frac{9}{2}$ and a finite electric quadrupole moment Q($^{73}$Ge) = -196 mb. In the bulk of the isotopically purified Ge quantum well, the diamond-cubic symmetry strongly suppresses the EFG and hence the quadrupolar splitting. Near the Ge/SiGe interface, however, strain, alloy disorder, and gate-induced electric fields break the local symmetry and generate finite EFGs. The residual $^{73}$e nuclei then form not only the conventional Larmor-linked hyperfine bath, but also a quadrupole-modified nuclear bath whose dynamics occur at EFG-set frequencies. Through the anisotropic hyperfine interaction, fluctuations of this quadrupolar bath are converted into qubit-frequency noise. This provides an additional dephasing channel: in Ramsey measurements it contributes to the low-frequency noise that limits $T_2^*$, while under Hahn-echo filtering it appears as finite-frequency spectral weight and produces extra collapse-and-revival features. The echo experiment therefore does not drive a conventional RF NQR transition; rather, the qubit senses this quadrupole-modified bath through its frequency-noise spectrum. We use the Zeeman-quadrupole Hamiltonian to identify the relevant nuclear branch frequencies. The single nuclear Hamiltonian in the EFG principal-axis frame is

$$H = H_Z + H_Q = -\gamma_{Ge-73}\hbar B \cdot I + \frac{eQV_{zz}}{4I(2I-1)}\left(3I_{z'}^2 - I(I+1) + \eta\left(I_{x'}^2 - I_{y'}^2\right)\right), \quad (1)$$

Here the primed axes define the local EFG principal-axis frame, $e$ is the electron charge, $V_{zz}$ is the principal EFG component and $\eta = (V_{xx} - V_{yy})/V_{zz}$ is the EFG asymmetry parameter. For an approximately axially symmetric EFG, the zero-field quadrupole term splits the ten nuclear spin projections into five degenerate Kramers pairs indexed by $|m|$. The adjacent-doublet transitions form an NQR-like quartet with frequencies in the ratio 1:2:3:4. With $\nu_Q = \frac{3eQV_{zz}}{2I(2I-1)h}$, the nominal zero-field branch frequencies are $\nu_Q, 2\nu_Q, 3\nu_Q, 4\nu_Q$.

Experimentally, the echo traces are fitted with a noise power spectral density (PSD) model containing charge noise, a conventional $^{73}$e Larmor-linked hyperfine contribution, and ensemble-broadened quadrupolar components, as described in the Methods and shown in Fig. 4c. The fit resolves a dominant field-stationary branch centred at 47.3 kHz, with a linewidth of order 3 kHz that reflects the distribution of local EFG environments, and another branch near 67–69 kHz. A separate field-scaling measurement at fixed field angle confirms that the 47.3 kHz branch remains stationary as $B$ is varied from 30 to 50 mT, whereas the Larmor-linked branch shifts linearly with $B$ (Extended Data Fig. 6). The field-stationary branch is close to the nominal $2\nu_Q$ quadrupolar branch, while the 67–69 kHz feature lies near both $\nu_L(B = 45\ \mathrm{mT}) \simeq 67.0$ kHz and the nominal $3\nu_Q$ branch. The time-domain dips in Fig. 4a should be interpreted as Hahn-echo filter maxima rather than as a one-to-one map to distinct transition frequencies. For a narrow spectral component at frequency $f$, the echo filter produces repeated maxima near $\tau_E \simeq (2n + 1)/f$. Thus, at $B \simeq 45$ mT, the 47.3 kHz component gives maxima near 21 and 63 μs, whereas the Larmor-linked/Zeeman-quadrupole mixed feature near 67 – 69 kHz gives maxima near 15, 44 and 73 μs. This accounts for the observed red, yellow, and blue arrow features without assigning each dip to a separate transition.

We use an angular dependent parameter $a_{\mathrm{Q}}$ to evaluate the effective qubit-visible spectral weight of the quadrupolar branch (Fig. 4f). Rotating the magnetic field changes the relative orientation among the Zeeman axis, the local EFG frame, and the anisotropic hyperfine-transduction tensor. Around 45 mT, the Zeeman and quadrupolar energy scales are comparable, so the finite-field mixed nuclear eigenstates can strongly modify the matrix elements sampled by the Hahn-echo filter. We capture the observed angular trend with a minimal signed-angle expansion (detailed in Method), which is consistent with quadrupolar tensor physics combined with qubit-dependent hyperfine transduction, without requiring a unique fixed EFG-axis geometry or a simple perturbative $\sin^2(2\theta)$ selection rule. Fig. 4f indicates that even a tiny angle rotation can significantly change the collapse visibility. A more complete understanding of the underlying NQR mechanisms may ultimately enable the use of hole spin qubits as sensors for high-resolution strain and electric-field-gradient mapping.

## Conclusion

In summary, high mobility isotopically purified Ge quantum well is synthesized by UHVCVD. These heterostructures enable the definition of high-performance, electrically gated quantum dots, serving as robust hosts for spin qubits. Due to the reduction of $^{73}$Ge abundance, $T_2^*$ of qubit reaches 22.3 μs on the hyperfine sweet spot, setting a new benchmark for hole qubit. $T_2^*$ of qubits off their sweet spots is enhanced to over 3 μs, consistent with the 20-fold reduction of $^{73}$Ge concentration. Single qubit fidelity over 99.9% is achieved both on and off sweet spot, proving both the necessity and the feasibility of isotopic purification as a pathway toward robust and scalable Ge spin-qubit architectures. Hahn-echo spectroscopy further reveals that, after the conventional Larmor-linked hyperfine response is reduced, residual $^{73}$Ge nuclei give rise to additional quadrupole-modified nuclear-noise components. The nearly field-stationary branch near 47.3 kHz and the finite-field mixed feature near 68.7 kHz are consistent with $^{73}$Ge nuclei sampling local EFGs generated by interface disorder, alloy disorder, strain and gate electrostatics. The observed response is not a direct RF-NQR measurement; rather, the hole qubit senses quadrupole-modified nuclear-bath dynamics through anisotropic hyperfine transduction into qubit-frequency noise. This quadrupolar channel represents an additional dephasing pathway, and its strong angular dependence suggests that local EFG environments can constrain the optimal operating angle in purified Ge devices. Conversely, the same sensitivity provides a spectroscopic handle on interfacial strain and EFG disorder, which may be useful for optimizing heterostructure growth and device design. Planar Ge with $^{73}$Ge concentration as low as 0.01% has been reported recently[51], this will not only further enhance the performance and robustness of hole qubits, but may also allow us to use single $^{73}$Ge nuclear spin as a new resource[52] for quantum computing.

# Acknowledgements

This work was supported by the National Key Research and Development Program (Grant No. 2024YFA1408200), and the National Natural Science Foundation of China (Grant No. 12374469, 12304099).

# Author Contributions

J. Zeng performed the experiments and analysed the data with the help from T. Pei. X. Tan performed the theoretical analysis. J. Lu grew the Ge/SiGe heterostructure. H. Wang and W. Bian fabricated the devices. Y. Zhang and J. Li characterize the Hall bar devices. C. Yang, L. Wu and Z. Guo helped with the measurement system setup. J.-W. Luo and Y. Wang reviewed and revised the article. T. Pei conceived and supervised the project. T. Pei, J. Zeng and X. Tan wrote the paper with inputs from all authors.

## Method

## Device Fabrication

Both the Hall bar device and the quantum dot device are fabricated on the same Ge/SiGe heterostructure with purified Ge quantum well. The native $SiO_2$ on Si cap is selectively removed by buffered oxide etch (BOE). Then Ohmic contacts are formed by depositing 40 nm Al layer followed by 300 °C annealing for 2 hours. 20 nm thick aluminium oxide ($Al_2O_3$) is grown by atomic layer deposition (ALD) as oxide layer. Subsequently, multilayer Al gate electrodes or Ti/Au gate electrodes can be deposited for quantum dot devices or Hall devices respectively.

## Measurement Setup

The quantum dot device is cooled in a dilution refrigerator (Bluefors LD series) with a base temperature of about 15mK. A 24-channel digital-to-analogue converter (QDevil QDAC) supplies voltages to the gate directly from the port at room temperature to the sample mounted on the homemade printed circuit board (PCB) at base temperature. The voltage pulse applied to the B2 and B4 are generated by an arbitrary waveform generator (Tektronix AWG5208). The EDSR microwave pulses applied to the B3 are generated by a vector signal generator (E8267D PSG). A lock-in Amplifier (Zurich Instruments UHFLI) outputs a 10 $mV_{pp}$ microwave to the gate $AG_{rf}$ and detect its reflected signal which is amplified by a cryogenic amplifier (ZW-LNA0.1-1B) at 4K plate. The vector magnet (American Magnetics AMI430) applies an external static magnetic field with an in-plane angle $\varphi_{\boldsymbol{B}} = 14^{\circ}$, which corresponds to the minimum in-plane effective g-factor for both qubits at $\theta_{\boldsymbol{B}} = 90.1^{\circ}$.

## Initialization, Manipulation and Readout

The pulse sequence scheme for the qubit control is shown in Fig. 2d. Each cycle consists of three main stages, initialization (I), manipulation (M) and readout (P, L), where the readout stage is constructed by two steps as shown in Fig. 2c. The asymmetric tunnelling rates from quantum dots to their neighbouring reservoirs have been pre-set, allowing us to use enhanced latching readout. The system is initialized to the S(2,0) states at I point for 10 μs through the exchange of holes with the reservoirs. The spins then rapidly reach point P and pause for 5 ns. Subsequently, in the main case, the spins pass adiabatically through the S(2,0)-$T_{-}$(1,1) anticrossing with a ramp time of 2 μs to arrive at the point M, where a state $T_{-}(1,1) = |\downarrow\downarrow\rangle$ is successfully prepared. The manipulation point M is sufficiently deep to ensure that the coupling between the two qubits is switched off. Following the coherent rotation of one of the qubits under the action of EDSR, the spins similarly return to point P with a ramp time of 2 μs and stop there for 50 ns. Finally, the spins rapidly ramp to L point in the occupation region (2, 1) and stay for 200 μs for readout.

## Randomized Benchmarking

In single-qubit RB, the native gates I, $\pi_x/2$ and $\pi_y/2$ compose the sequences of Clifford gates[21]. At the end of each sequence, a recovery rotation is applied to bring the qubit back to its initial state, that is $|\downarrow\downarrow\rangle$ in our case. Experimentally, the gate I is compiled into an idle operation for 5 ns. There is a 5 ns waiting time between each native gate. For each sequence length, 30 random combinations of Clifford gates are sampled. For each combination, the parallel spin probability is defined as the average result from 1000 single-shot measurement repetitions. The experimental data are fitted using $P(N) = A_0\,(2F_{avg}-1)^N + B_0$, where $F_{avg}$ is the average Clifford fidelity and $N$ is the number of Clifford gates in the sequence. All errors related to state preparation or measurement (called SPAM errors) are absorbed in the constants $A_0$ and $B_0$. Considering the average 3.625 generators per Clifford composition, the average gate fidelity is given by $F_g = 1 - (1 - F_{avg})/3.625$. The uncertainties of the fidelity in the main text represent 1 s.d. acquired from the curve fitting.

## Zeeman-Quadrupole Mixing Model

The relevant $^{73}$Ge nuclei couple to both the external magnetic field and the local EFG tensor. For nucleus $j$, in the EFG principal-axis frame ($x'$, $y'$, $z'$), when a magnetic field is applied, the full Hamiltonian (M1) is numerically diagonalized in the $|I, m\rangle$ basis of the EFG frame. Writing $\boldsymbol{B} = (B_{x',j}, B_{y',j}, B_{z',j})$, the $j$ th nuclear Hamiltonian is

$$H_j = H_{Z,j} + H_{Q,j} = -\hbar\gamma_{Ge-73}\left(B_{x',j}I_{x',j} + B_{y',j}I_{y',j} + B_{z',j}I_{z',j}\right)$$

$$+\frac{eQV_{zz,j}}{4I(2I-1)}\left(3I_{z',j}^2 - I(I+1) + \eta_j\left(I_{x',j}^2 - I_{y',j}^2\right)\right), \quad \text{(M1)}$$

where $Q$ is the electric quadrupole moment of $^{73}$Ge, $V_{zz,j}$ is the principal EFG component, $\eta_j = \left(V_{xx,j} - V_{yy,j}\right)/V_{zz,j}$ is the local EFG asymmetry. For an axially symmetric EFG ( $\eta_j = 0$ and $B = 0$ ), the zero-field eigenvalues are $E_m^{(0)} = \frac{3eV_{zz}Q}{4I(2I-1)}\left(m^2 - \frac{I(I+1)}{3}\right)$, and the allowed $\Delta m = \pm 1$ transitions form a quartet with frequencies $\nu_{Q,j}$, $2\nu_{Q,j}$, $3\nu_{Q,j}$ and $4\nu_{Q,j}$ in the exact ratio 1:2:3:4, where $\nu_{Q,j} = \frac{3eQV_{zz,j}}{2I(2I-1)h}$ is the fundamental quadrupole frequency. For $I = \frac{9}{2}$, this reduces to $\nu_{Q,j} = \frac{eQV_{zz,j}}{24h}$. At finite magnetic field, the Zeeman and quadrupolar terms are generally not diagonal in the same basis $|I,m\rangle$. The nuclear eigenstates of the finite-field nuclear Zeeman-quadrupole Hamiltonian $H_{\text{nuc}}|\psi_a\rangle = E_a|\psi_a\rangle$, should therefore be written as hybridized eigenstates in the EFG-frame basis, $|\psi_a\rangle = \sum_{m=-I}^{I} c_{am}|I,m\rangle$,where $\sum_m |c_{am}|^2 = 1$, rather than as a single pure $|m\rangle$ projection. The longitudinal Zeeman component $B_{z',j}$ shifts the axial $|m\rangle$ levels, whereas the transverse components $B_{x',j}I_{x',j} + B_{y',j}I_{y',j}$ couple $\Delta m = \pm 1$ states. The quadrupolar asymmetry term

$$\eta_j\left(I_{x'}^2 - I_{y'}^2\right) = \left(\eta_j/2\right)\left(I_+^2 + I_-^2\right), \quad \text{(M2)}$$

couples $\Delta m = \pm 2$ states, where $I_\pm = I_x \pm iI_y$ are the nuclear-spin raising and lowering operators in the local EFG frame. In the intermediate-coupling regime $\nu_L \approx n\nu_Q$, the observed branches are therefore obtained from eigenvalue differences of the full Zeeman-quadrupole Hamiltonian, with $|m\rangle$ labels used only as adiabatic references to the zero-field quadrupolar branches.

The hole spin couples to the residual nuclear dynamics through the anisotropic dipole-dipole hyperfine interaction, while the contact hyperfine term is strongly suppressed for p-like valence-band states. With the dipole-dipole hyperfine tensor $\mathsf{A}_k$ we write this coupling as

$$H_{\text{hf}} = \sum_k \mathbf{S} \cdot \mathsf{A}_k \cdot \mathbf{I}_k. \quad \text{(M3)}$$

After projection onto the qubit quantization axis $\mathbf{n}_q$, nuclear-spin fluctuations produce a qubit-frequency shift,

$$\delta\omega_q(t) = \frac{1}{\hbar}\sum_k \mathbf{n}_q \cdot \mathbf{A}_k \cdot \delta\mathbf{I}_k(t). \quad \text{(M4)}$$

Thus the echo signal is most sensitive to nuclear-spin transitions that modulate the Overhauser field along $\mathbf{n}_q$. In this picture, the visibility of a quadrupole-modified branch is controlled not only by its transition angular frequency $\omega_{ab} = \frac{|E_a - E_b|}{\hbar}$, but also by an effective transduction matrix element between the mixed nuclear eigenstates,

$$S_{\delta\omega}(\omega) \propto \sum_k \sum_{a,b} p_{k,a} \left|\langle\psi_{k,a}|\mathbf{n}_q \cdot \mathbf{A}_{\boldsymbol{k}} \cdot \mathbf{I}|\psi_{k,b}\rangle\right|^2 \exp\left[-\frac{\left(\omega - \omega_{k,ab}\right)^2}{2\sigma^2}\right], \quad \text{(M5)}$$

$$\omega_{k,ab}(B,\theta_B) = \frac{\left|E_{k,a}(B,\theta_B) - E_{k,b}(B,\theta_B)\right|}{\hbar}.$$

This keeps the leading hyperfine-transduction channel by which quadrupole-modified nuclear dynamics are converted into qubit-frequency noise. Higher-order electric pathways, such as nuclear-quadrupole-induced orbital or g-tensor modulation followed by spin-orbit transduction, are not resolved independently by the present echo data. The finite-frequency part of the qubit frequency noise is modelled as arising from nuclear-spin transitions that modulate the Overhauser field along the hole quantization axis. In the Hahn-echo experiment, the decay function is[53]

$$P(\tau) = b + c\exp(-\chi(\tau)), \quad \text{(M6)}$$

with the Hahn-echo filter function

$$F(\omega,\tau) = \frac{8\sin^4\left(\frac{\omega\tau}{4}\right)}{\omega^2}, \tag{M7}$$

and the spectral integral $\chi(\tau) = \frac{1}{\pi}\int_0^\infty S(\omega)F(\omega,\tau)d\omega$. For a narrow spectral component at ordinary frequency $f = \omega/2\pi$, the Hahn-echo filter produces maxima approximately at $\tau_E \simeq \frac{2n+1}{f}, n = 0,1,2, \ldots$

Thus repeated dips in the echo traces can arise from the same spectral component sampled at different filter orders. The noise spectrum is modelled as

$$S(\omega) = \frac{S_{0,\mathrm{E}}}{|\omega|} + S_{0,\mathrm{HF}}\exp\left[-\frac{(\omega-\omega_L)^2}{2\sigma_{\mathrm{HF}}^2}\right] + a_Q S_{0,\mathrm{HF}} \sum_i w_i \exp\left[-\frac{(\omega-\omega_i)^2}{2\sigma_{\mathrm{NQR}}^2}\right]. \tag{M8}$$

where the first term describes charge noise, the second term is the conventional Larmor-linked [73]Ge hyperfine contribution with Larmor frequency $\omega_L$, and the third term describes the effective NQR-mediated nuclear-noise component. Here $a_Q$ is a dimensionless effective NQR visibility normalized to the hyperfine noise amplitude of the same trace, and $w_i$ are global branch weights satisfying $\sum_i w_i = 1$. Thus the dimensional quadrupolar spectral weight extracted by the fit varies with angle $\theta_B$ is $a_Q(\theta_B)S_{0,\mathrm{HF}}(\theta_B)$. We normalize these branch weights to the hyperfine noise amplitude of the same trace by writing $A_i(\theta_B) = S_{0,\mathrm{HF}}(\theta_B)\, a_Q(\theta_B)\, w_i$, this gives the effective quadrupolar term used in the echo fits,

$$S_Q(\omega,\theta_B) = a_Q(\theta_B)S_{0,\mathrm{HF}}(\theta_B) \sum_i w_i \exp\left[-\frac{(\omega-\omega_i)^2}{2\sigma_{\mathrm{NQR}}^2}\right]. \tag{M9}$$

Equivalently, the fitted visibility can be viewed as $a_Q(\theta_B) = \frac{1}{S_{0,\mathrm{HF}}(\theta_B)}\sum_i A_i(\theta_B)$. Since $a_Q$ is a positive spectral-weight parameter, a direct polynomial fit to $a_Q$ is not ideal because it does not guarantee positive values. Moreover, the observed quadrupolar visibility is an effective quantity set by several multiplicative factors, including the quadrupolar transition matrix elements, the local EFG ensemble, the hyperfine-transduction efficiency, and the Hahn-echo filter response. We therefore parameterize its angular dependence through $g(\theta_B) = \log a_Q(\theta_B)$, for which multiplicative variations in $a_Q$ become additive contributions to $g$. If the angular dependence is smooth in the narrow range around a reference angle $\phi^*$ close to the hyperfine sweet spot, $g(\theta_B)$ can be expanded locally as a Taylor series,

$$g(\theta_B) = g(\phi^*) + g'(\phi^*)(\theta_B - \phi^*) + \frac{1}{2}g''(\phi^*)(\theta_B - \phi^*)^2 + \cdots$$

Keeping terms up to second order and defining the dimensionless signed angle, with $a_Q^* = a_Q(\phi^*)$ we write

$$g(\theta_B) \simeq \log a_Q^* - \kappa_1 x_B - \kappa_2 x_B^2, \tag{M10}$$

$$a_Q(\theta_B) = a_Q^* \exp(-\kappa_1 x_B - \kappa_2 x_B^2), \tag{M11}$$

where $x_B = \frac{\theta_B - \phi^*}{1^\circ}$, $a_Q^* = 3.87, \kappa_1 = 3.09, \kappa_2 = 2.93$, $\phi^* = 90.5^\circ$ the reference angle close to the hyperfine sweet spot. The NQR branch centres $\omega_i$ are treated as global parameters. In the four-branch model they correspond to the NQR-like multiplet of interfacial $^{73}$Ge, with the dominant branch near $\omega_i/2\pi \simeq 47.3$ kHz. The linewidth $\sigma_{\mathrm{NQR}}/2\pi \simeq 3$ kHz captures the effective inhomogeneous broadening of the interfacial EFG distribution.

Following the anisotropic-hyperfine description of Ge hole qubits, we constrain the conventional hyperfine background independently using the measured g-tensor anisotropy and Ramsey $T_2^*$ data (Extended Data Fig. 5). The hyperfine projection is $P_{\mathrm{HF}}(\theta_{\mathrm{B}}) = \left|\mathbf{n}_q(\theta_{\mathrm{B}}) \cdot \mathbf{n}_{\mathrm{HF}}\right|^2$, where $\mathbf{n}_{\mathrm{HF}}$ is the effective direction of the conventional hyperfine Overhauser-field fluctuations, and the boundary curve $\frac{1}{T_2^{*2}(\theta_B)} = \Gamma_0 + \Gamma_{\mathrm{HF}} P_{\mathrm{HF}}(\theta_B)$ is used to fix the scale of $S_{0,\mathrm{HF}}$ before fitting the NQR component. Here $\Gamma_0$ is the angle-independent baseline contribution to $1/T_2^{*2}$, and $\Gamma_{\mathrm{HF}}$ sets the projected conventional-hyperfine dephasing scale; both have units of inverse time squared. From the fitted $\nu_Q = 23.54$ kHz should be viewed as an ensemble-effective axial quadrupole scale for the qubit-visible $^{73}$Ge nuclei. It corresponds to the quadrupole coupling constant is $C_Q = \frac{eV_{zz}Q}{h} = \frac{2I(2I-1)\nu_Q}{3} = 565$ kHz, consistent with strain-induced EFGs at semiconductor interfaces[38,54].

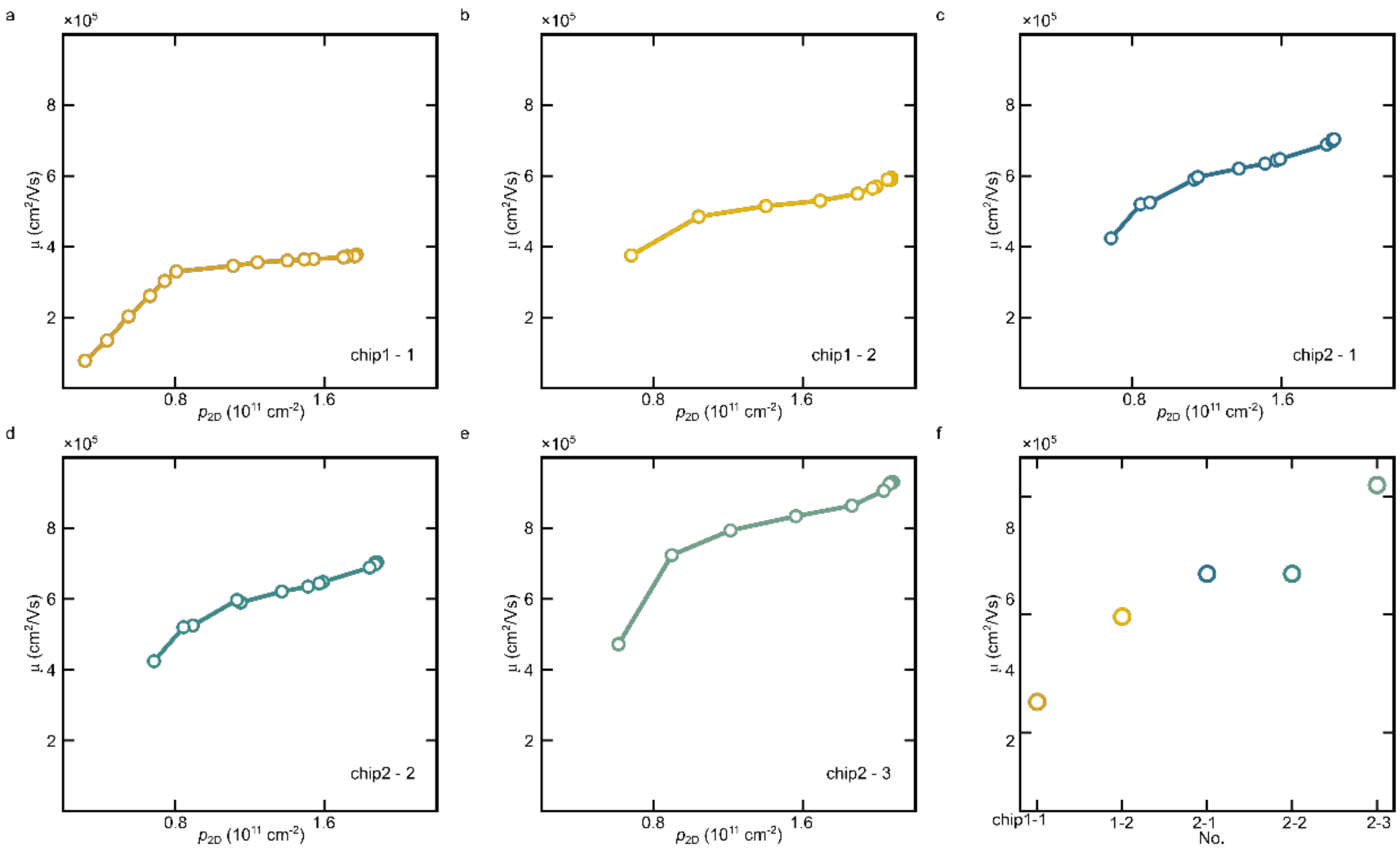

**Extended Data Fig. 1 Density dependence of hole mobility for different Hall devices from the same isotopically purified wafers.** The wafer is diced into 1 cm × 1 cm chip. Hall devices are fabricated on two chips. Two devices are tested from chip 1 (**a, b**) and three from chip 2 (**c-e**). **f,** The maximum mobilities obtained for the five devices. Device 1-1 is characterized in the full density range and shown in the main text. Quantum dot devices are fabricated on other chips from the same wafer.

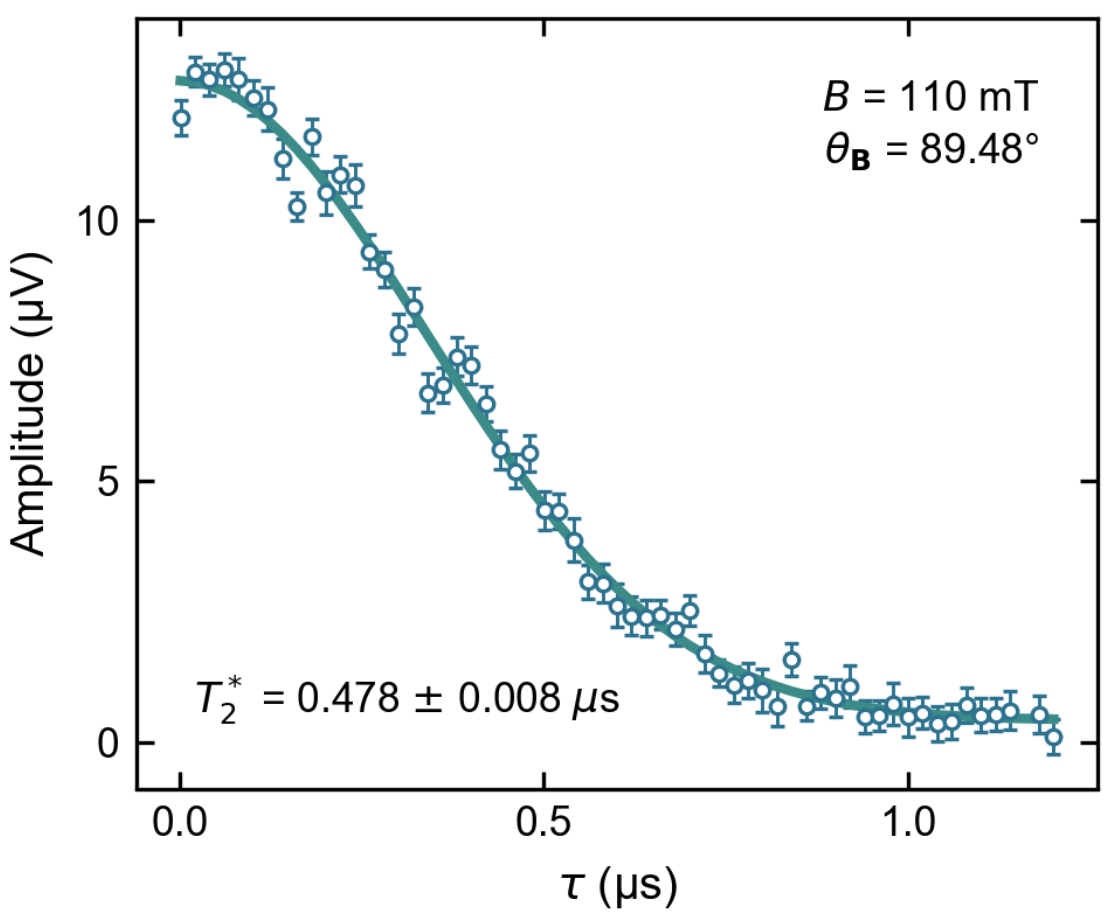

**Extended Data Fig. 2 Dephasing time of a natural planar Ge/SiGe device fabricated with similar recipes and measured in the same measurement setup.** The dephasing time $T_2^*$ are measure by Ramsey experiment at $\theta_B$ = 89.48° at $B$ = 110 mT, and remains an order of magnitude lower than that of the isotopically purified sample discussed in the main text.

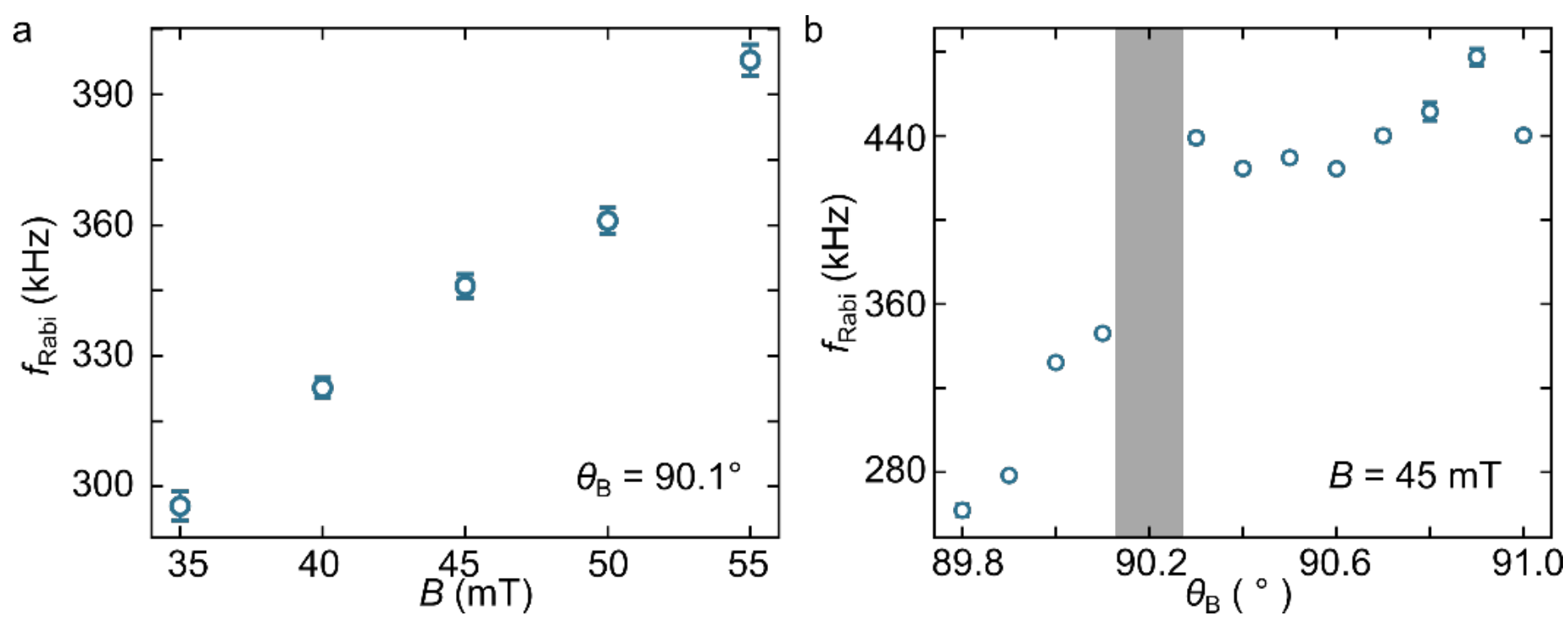

**Extended Data Fig. 3. Magnetic field dependence of qubit's (Q1) Rabi frequency. a,** Rabi frequency as a function of field magnitude. The linear dependence is consistent with the $g$-tensor modulation model. **b,** Angular dependence of the Rabi frequency, which reaches its minimum on the sweet spot. The dark zone corresponds to the regime where two qubits' Larmor frequencies are close to each other so that are unable to be distinguished.

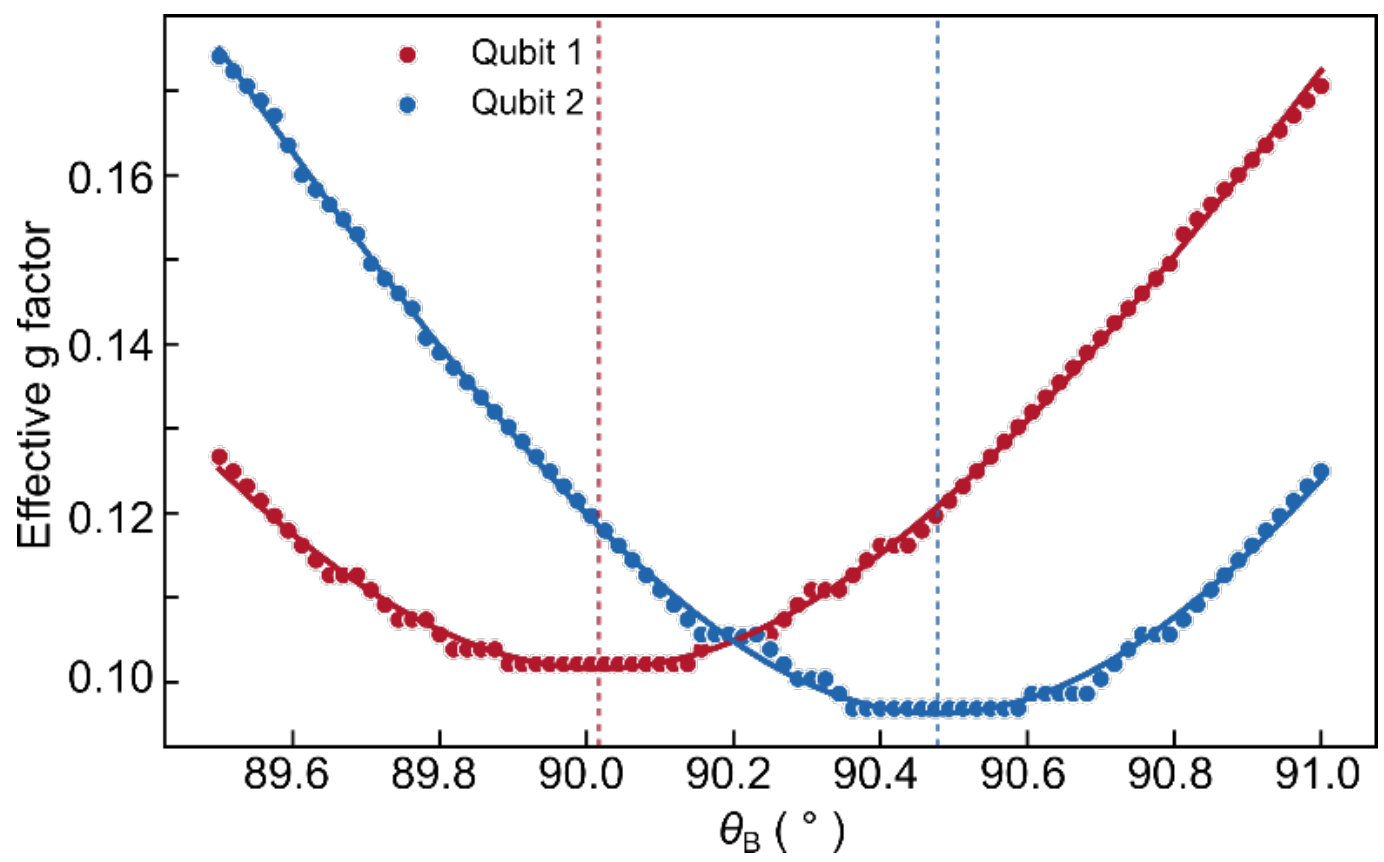

**Extended Data Fig. 4.** Effective g-factor of the two qubits as a function of the polar angle around the hyperfine sweet plane, extracted from data in Fig. 3b in the main text. The effective g-factor is a combination of in-plane g factor and out-of-plane g factor.

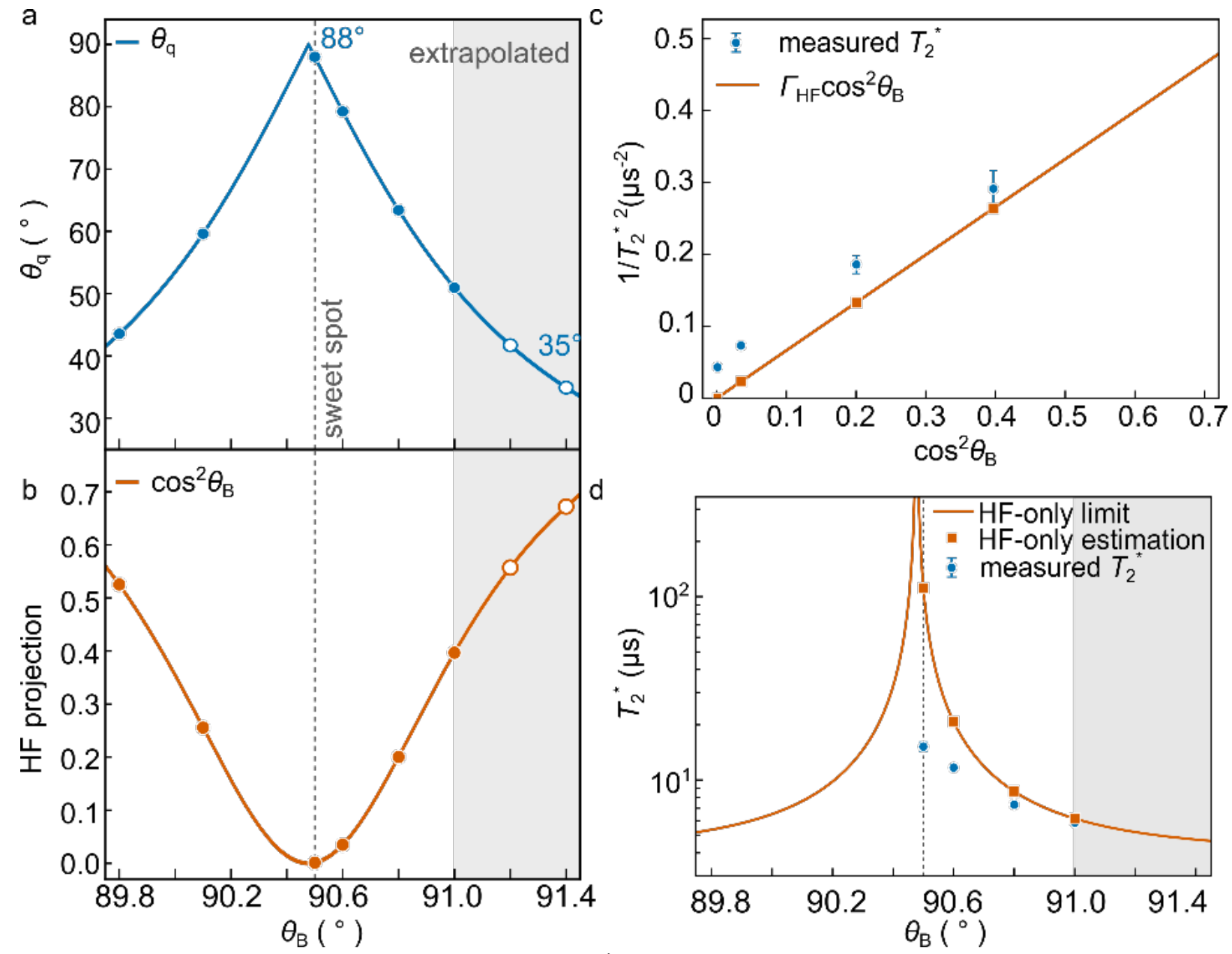

**Extended Data Fig. 5 | Hyperfine-limited $T_2^*$ envelope inferred from the Q2 local $g$-factor anisotropy.** **a,** Q2 quantization-axis angle $\theta_B$, converted from the extracted in-plane and out-of-plane g-factor components. The grey shaded region denotes the angular range where $T_2^*$ is extrapolated beyond the measured Ramsey data. **b,** Ising-type hole-nuclear hyperfine projection, $P_{HF}(\theta_B) = |\mathbf{n}_q(\theta_B) \cdot \mathbf{n}_{HF}|^2$, calculated from the quantization-axis map. The markers in **a** and **b** indicate the field angles used for the measurements in Fig. 4 of the main text. **c,** Squared Ramsey dephasing rate, $1/(T_2^*)^2$, plotted against the hyperfine-projection factor. The orange line is a hyperfine-boundary fit including an angle-independent dephasing floor near the sweet spot, and is used to estimate $\Gamma_{HF}$. **d,** Hyperfine-limited $T_2^*$ envelope obtained by mapping the fitted $\Gamma_{HF}$ boundary back to field angle. The orange curve gives the estimated conventional quasi-static hyperfine limit, while the grey shaded region denotes extrapolation outside the measured Ramsey range. This estimate constrains the angular scale of the conventional hyperfine contribution used in Fig. 4, but is not a fit to the finite-frequency quadrupolar-mediated collapse feature.

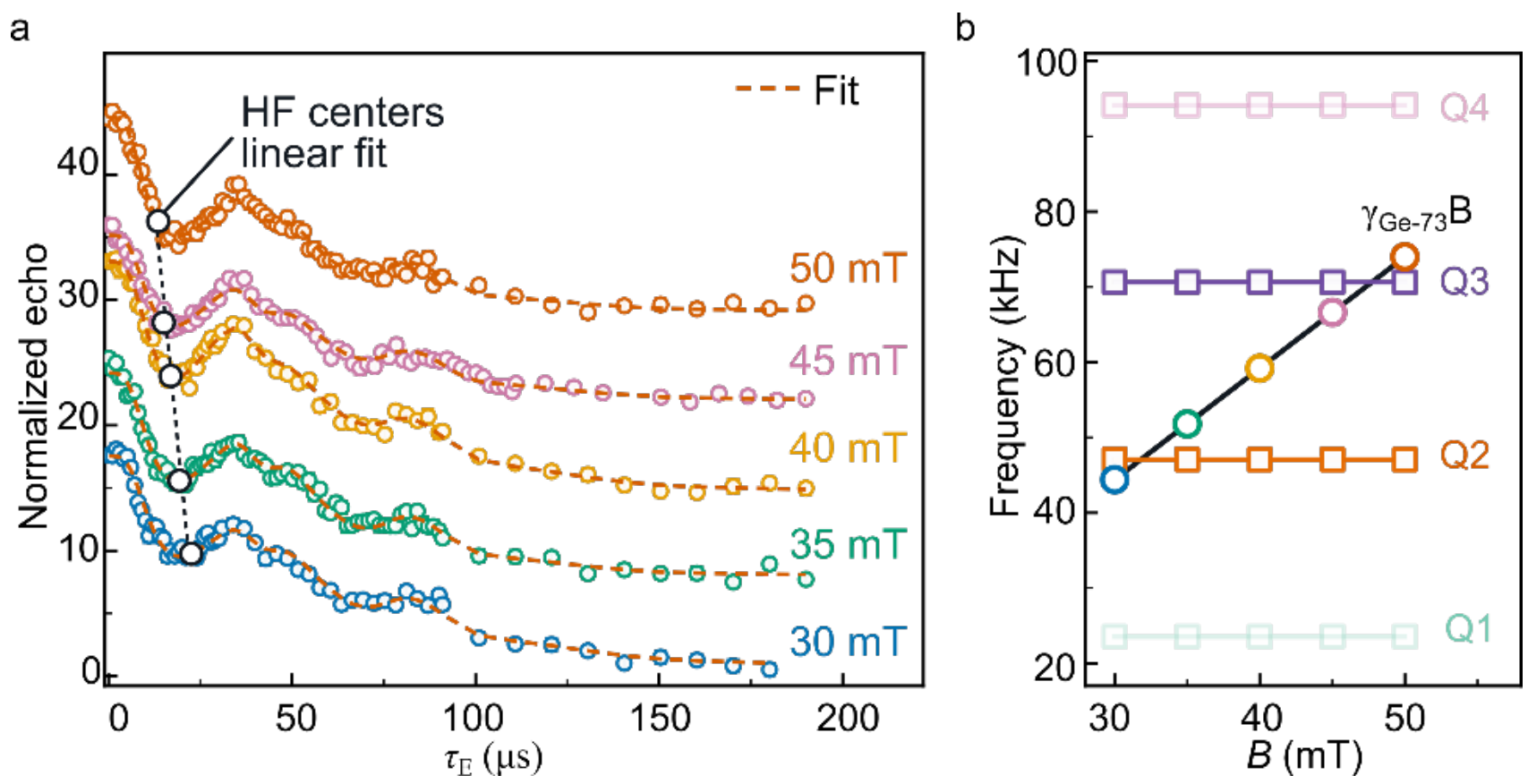

**Extended Data Fig. 6 Field-scaling diagnostic of Larmor-linked hyperfine and field-stationary quadrupolar spectral components. a,** Hahn-echo traces of Q1 measured at $\theta_B = 90.1°$ while varying the magnetic field from 30 to 50 mT. The traces are vertically offset; points are color-coded by field, and dashed curves are fits using charge noise, a $^{73}$Ge Larmor-linked hyperfine term, and one approximately field-independent NQR component at $\nu_2 = 2\nu_Q = 47.1$ kHz. The marked dip centres follow the expected field-dependent timing of the Larmor-linked feature. **b,** Frequency-branch analysis. The $^{73}$Ge hyperfine branch shifts from 44.7 kHz to 74.5 kHz, consistent with Larmor scaling, whereas the $\nu_2$ quadrupolar branch remains stationary at 47.3 kHz within experimental resolution. The opacity of the horizontal $Q_i$ branches indicate the fitted relative branch weight $w_i$. At 30 mT, the Larmor branch crosses $\nu_{2,3}$, producing the interference visible in **a**. The stationary branch is assigned to the nominal $\pm 3/2 \leftrightarrow \pm 5/2$ quadrupolar branch, supported by the field-scaling analysis and by the angle-dependent effective visibility $a_Q$ in Fig. 4f.